\documentclass{appolb}

\usepackage{epsfig}
\usepackage{amsmath}
\usepackage{amssymb}
\usepackage{color}
\usepackage{graphicx}
\usepackage{hyperref}

\begin{document}
\title{
On the 
history of multi--particle production \\ in high energy collisions %
\thanks{Presented at Strangeness in Quark Matter 2011, Krakow, Poland   }%
}
\author{Marek Gazdzicki
\address{Institut fur Kernphysik, Goethe-University, 
         Max--von--Laue-Str.~1
         60438 Frankfurt am Main, Germany and \\ 
         Instytut Fizyki, Jan Kochanowski University, 
         ul. Swietokrzyska~15, 25--406 Kielce, Poland}
}
\maketitle
\begin{abstract}
The 60th birthday of Johann Rafelski was celebrated
during the Strange -ness in Quark Matter 2011 in Krakow.
Johann was born in Krakow and he initiated
the series of the SQM conferences.
This report, which briefly presents my personal view on a history
of multi--particle production in high energy collisions, is dedicated to Johann.
\end{abstract}
  
\section{Introduction}
The systematic study of particle production in collisions of high energy
particles started about 60 years ago with the construction of first
accelerators. Here, I will briefly present my personal view on the history of
this era. Johann Rafelski, whose 60th birthday was celebrated
at the Strangeness in Quark Matter conference in Krakow has been one of the
key contributors since the mid-70s~\cite{Raf1,Raf2}. 
Therefore
his impact on the  development of
the field will be emphasized. 

\newpage
\section{Experimental and theoretical status quo}

\begin{figure}[!h]
\begin{center}
\includegraphics[width=0.8\columnwidth]{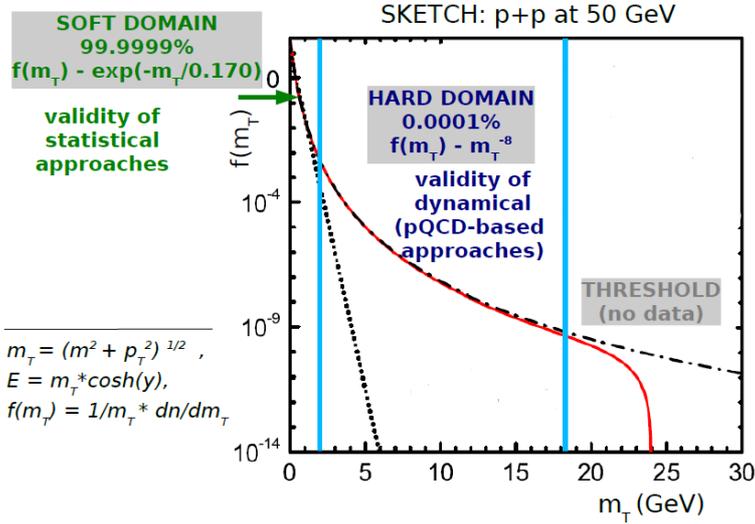}
\end{center}
  \caption{\label{fig:status-quo} 
Experimental and theoretical status quo of multi--particle production
in high energy collisions.
}
\end{figure}

 The experimental and theoretical status quo of multi--particle
 production in high energy collisions is summarized in 
 Fig.~\ref{fig:status-quo}, where a sketch of
 the transverse mass, $m_T$, spectra of hadrons produced in p+p 
 interactions at the center--of--mass energy $\sqrt{s} = 50$~GeV
 is shown~\cite{Begun:2008fm}.
 Clearly, there are three distinct domains:

 \begin{enumerate}

 \vspace*{-0.2cm}
 \item 
  {\it the soft domain}, $m_T \lesssim 2$~GeV, in which a predominant
  majority of all particles is produced, 
  $m_T$ spectra are exponential and produced particles are essentially
  uncorrelated; 
  their production properties are well described by statistical and 
  hydrodynamical models,

 \vspace*{-0.2cm}
 \item 
  {\it the hard domain}, $2 \lesssim m_T < \sqrt{s}/2$~GeV, 
  in which only a  small
  fraction of all produced particles is located,   
  $m_T$ spectra follow a  power--law dependence 
   and produced particles are strongly correlated (organized in so--called jets);
  in this domain
  particle production properties are best described by dynamical QCD--based
  approaches,

 \vspace*{-0.2cm}
 \item 
 {\it the threshold domain}, $m_T \approx \sqrt{s}/2$, 
 the spectrum in this domain is not measured due to a very low particle yield,  
 it is believed to steeply decrease to zero with $m_T$ approaching
 its threshold value given 
 by energy--momentum conservation laws, $\sqrt{s}/2$.

 \end{enumerate}

\newpage
\section{History of multi--particle production in short}

\begin{figure}[!h]
\begin{center}
\includegraphics[width=0.8\columnwidth]{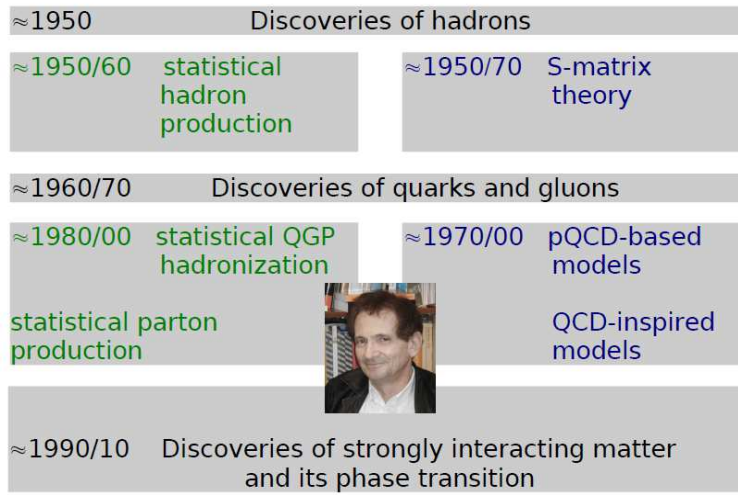}
\end{center}
  \caption{\label{fig:history} 
History of multi--particle production in short.
Johann Rafelski's contributions span between statistical and
QCD--inspired approaches and are deeply rooted in experiment.
}
\end{figure}

A brief history of multi--particle production is shown in 
Fig.~\ref{fig:history}. The field started in the 50s with discoveries
of hadrons, first in cosmic--ray experiments and soon after
in experiments using beams of particles produced in accelerators.
Two classes of models were developed in this time, namely
statistical and dynamical models of hadron production.
The latter class was initiated by the scattering matrix approach.
Systematic results on properties of hadrons and their interactions with electrons
accumulated in the 60-70s led to discoveries of sub--hadronic particles, quarks and gluons.
The idea of the quark--gluon plasma (QGP) was formulated.
Subsequently,
statistical approaches of 
quark and gluon hadronization and a statistical model of quark and
gluon production were developed.
In parallel, a dynamical theory of strong interactions (QCD) was
established. Many QCD--based and QCD--inspired models of multi--particle
production appeared. 
Finally, experimental studies of nucleus--nucleus collisions at
high energies resulted in discoveries of strongly interacting matter
and its phase transition between its hadron gas and quark--gluon plasma
forms. 

\emph{ 
Johann Rafelski 
greatly contributed  to the field  by developing both statistical and
QCD--inspired approaches to the quark--gluon plasma and its hadronization.
For a long time his ideas inspired experimental developments and he was always
devoted to the interpretation of new data.
}

\newpage
\section{Discoveries of hadrons}

\begin{figure}[!h]
\begin{center}
\includegraphics[width=1.0\columnwidth]{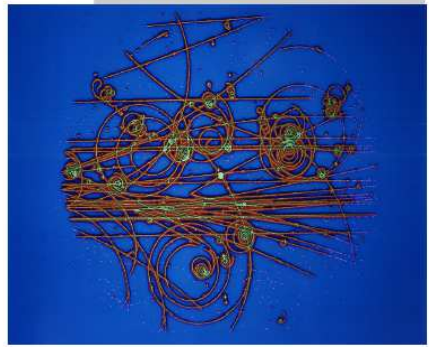}
\end{center}
  \caption{\label{fig:hadrons} 
Discoveries of hadrons and a brief history of particle
accelerators.
The maximum beam energy of accelerators is given in the fixed
target system. Accelerators used so far in the study of multi--particle
production are indicated in red.
}
\end{figure}

Naturally, the first hadrons, discovered in collisions of 
cosmic-ray particles, were the lightest ones, pion,  kaon and $\Lambda$.
With the rapid advent of particle accelerators (see Fig.~\ref{fig:hadrons}
for a brief history)
new particles were uncovered almost day--by--day.
There are about 1000 hadronic states known so far. 
Their density in mass  increase approximately 
exponentially~\cite{Broniowski:2004yh}
as predicted by  Hagedorn's Statistical 
Bootstrap Model~\cite{Hagedorn:1965st}.

\newpage
\section{Statistical hadron production}

The first statistical model of multi--hadron production
was proposed by Fermi~\cite{Fermi:1950jd}. 
He  assumed that hadrons produced
in high energy collisions are in equilibrium and that the
energy density  of the created hadronic system increases with
increasing collision energy.
Soon after, 
Pomeranchuk~\cite{Pomeranchuk:1951ey}
pointed out that hadrons cannot decouple (freeze--out)
at high energy densities. They will rather
continue to interact while expanding until the matter density is low
enough for interactions to be neglected. He estimated the 
freeze-out temperature to be close to pion mass, $\approx$150~MeV.
Inspired by this idea Landau~\cite{Landau:1953gs},
and his collaborators formulated a quantitative
hydrodynamical model describing the expansion of  strongly interacting
hadronic matter between  Fermi's equilibrium high density stage (the early stage)
and  Pomeranchuk's low density decoupling stage (the freeze--out).
The Fermi--Pomeranchuk--Landau picture serves as a base for modeling
high energy nuclear collisions up to now~\cite{Florkowski_textbook}.

In the 60s Hagedorn made  an important conjecture, namely that
matter composed of hadrons has a maximum temperature, 
the so--called Hagedorn temperature $T_H \approx 150$~MeV~\cite{Hagedorn:1980cv}.
This conjecture was based on his Statistical Bootstrap Model.
Note, that it was in  contradiction to 
the Fermi's model in which the temperature of hadronic matter
created at the early stage of collisions increases monotonically
with collision energy and it is unlimited.

\begin{figure}[!h]
\begin{center}
\includegraphics[width=0.7\columnwidth]{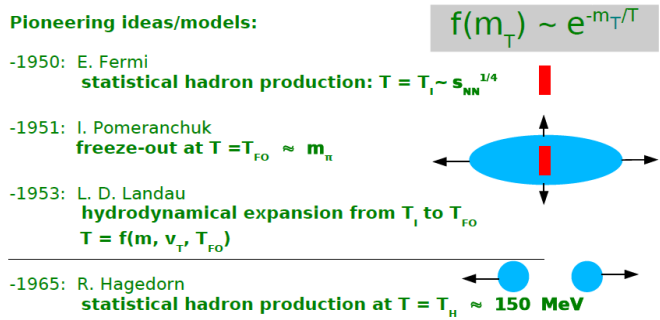}
\end{center}
  \caption{\label{fig:statistical} 
Pioneering models and ideas on statistical
hadron production.
}
\end{figure}

Figure~\ref{fig:statistical} sketches a brief history of
pioneering ideas and models on statistical hadron production.

The statistical and hydrodynamical models predict an
approximately exponential form of particle  
transverse mass spectra,
provided collective flow of matter developed in the course of
expansion is small. 

\newpage
\section{S-matrix theory}

The scattering (S--)matrix theory,  which relates the initial state and the final state of  
a scattering process, was initiated by Wheeler and Heisenberg in the 30--40s.  
It led to the development of the 
Regge theory studying
analytic properties of the scattering amplitude treated as a function of
angular momentum,
and it finally resulted in the string theory
pioneered by Veneziano.
These theories, formulated without specifying elementary particles, 
do not refer to a
space--time structure of interaction processes. They assume that
all particles are bound states lying on Regge trajectories 
and scatter self-consistently. 

\begin{figure}[!h]
\begin{center}
\includegraphics[width=0.65\columnwidth]{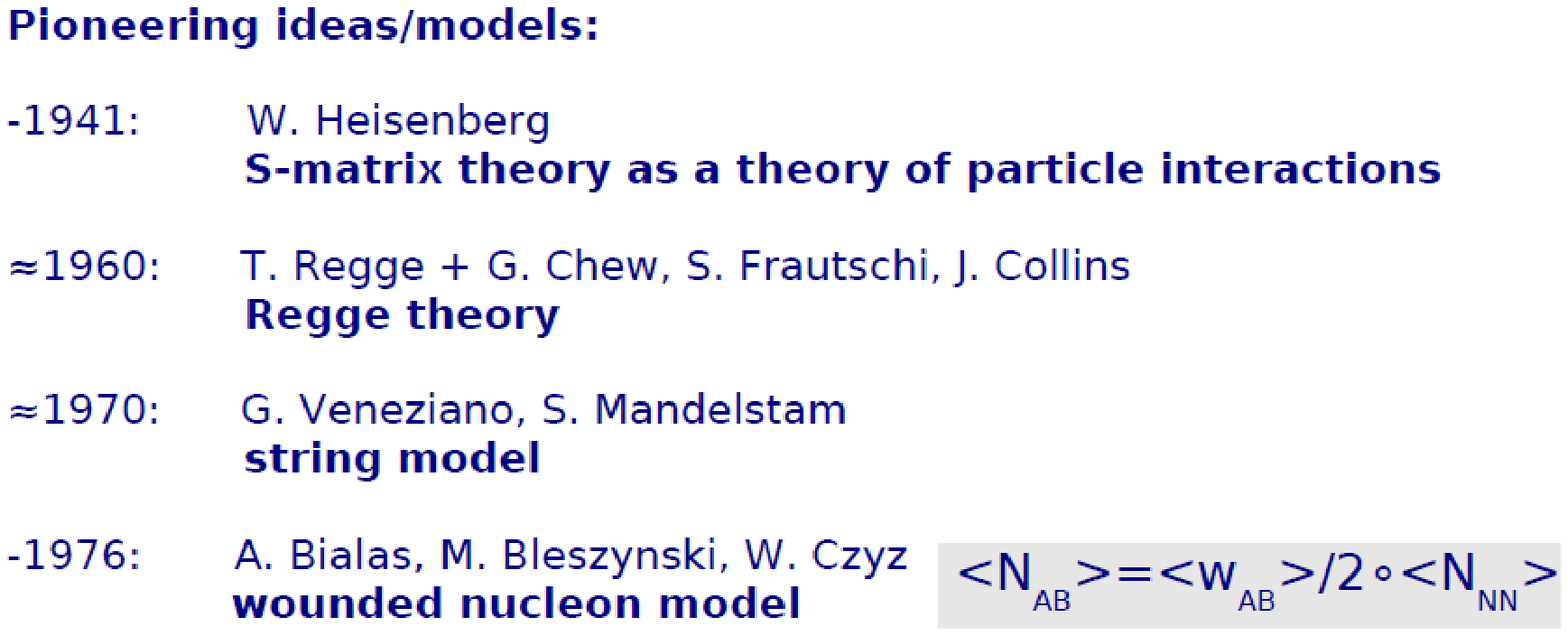}
\end{center}
\vspace*{-0.2cm}
  \caption{\label{fig:smatrix} 
Milestones of the S-matrix era.
}
\end{figure}

Within the spirit of these theories the 
Wounded Nucleon Model (WNM) was proposed 
by Bialas, Bleszynski and Czyz~\cite{Bialas:1976ed} 
in 1976.
It assumes that particle production in nucleon--nucleus and nucleus--nucleus
collisions is an incoherent superposition of particle production from
wounded nucleons, i.e. nucleons which interacted inelastically and whose
number is calculated using the Glauber approach.
Up to now, predictions of the WNM model have remained an
important baseline for interpretation of experimental data.
In particular, in the case of mean hadron multiplicities the WNM predicts:
\begin{equation}
\langle n_{AB} \rangle = \langle w_{AB} \rangle/2 \cdot \langle n_{NN} \rangle ,
\end{equation}   
where $\langle n_{AB} \rangle$ and $\langle n_{NN} \rangle$ are
mean hadron multiplicities in A+B collisions and nucleon-nucleon
interactions, respectively, whereas $ w_{AB} $ is a mean number
of wounded nucleons in A+B collisions.
Note, that the above WNM prediction resembles the corresponding
prediction of statistical models in the grand canonical approximation
(thermodynamical models) providing the number of wounded nucleons is
replaced by a system volume. This explains an approximate validity 
of the WNM for the yield of pions, the most popular hadrons.
However, the model significantly fails already for yields of strange 
hadrons. 
Figure~\ref{fig:smatrix} presents milestones of the S-matrix era.

\newpage
\section{Discoveries of quarks and gluons}

\begin{figure}[!h]
\begin{center}
\includegraphics[width=1.0\columnwidth]{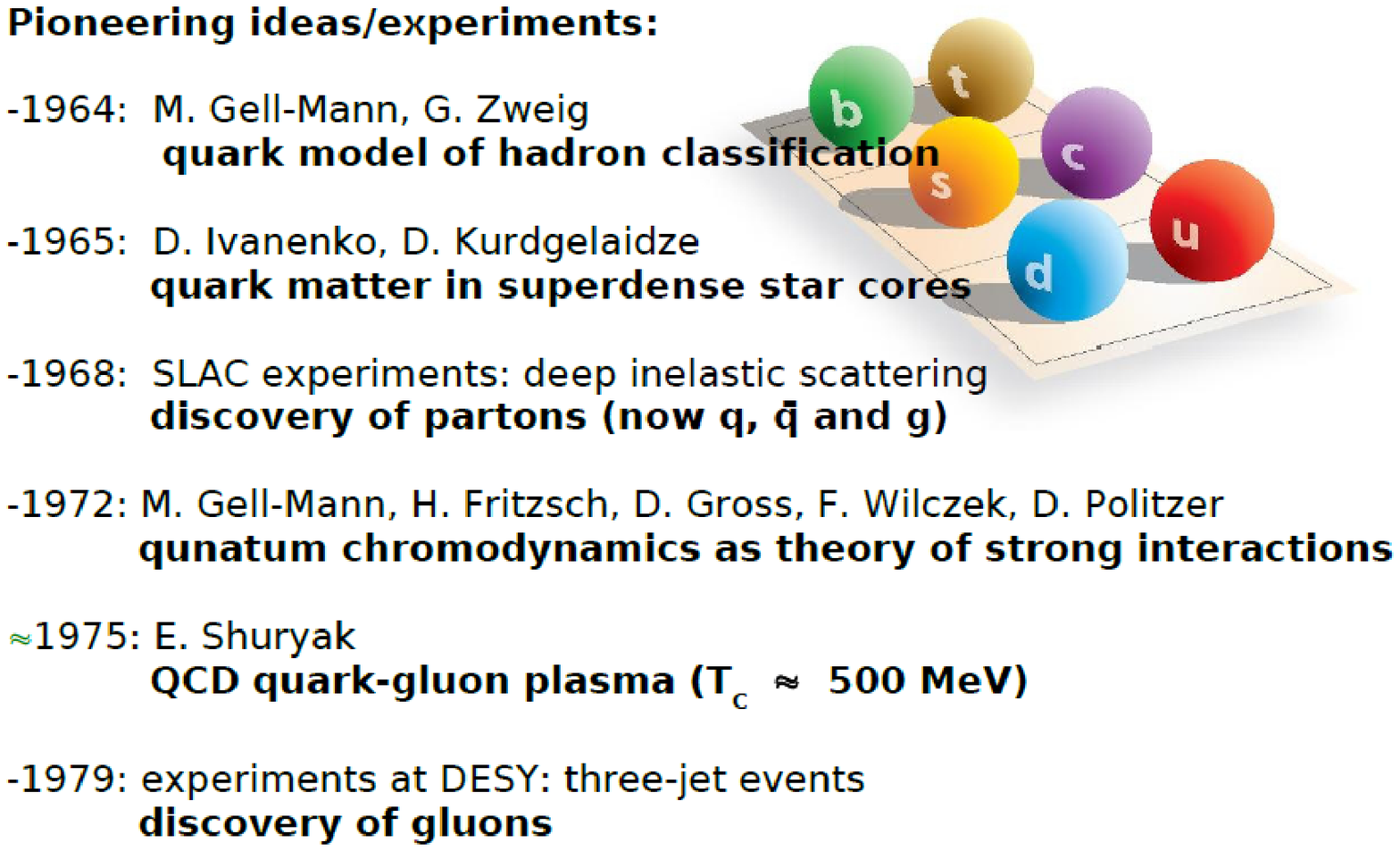}
\end{center}
  \caption{\label{fig:quarks-gluons}
A brief history of discoveries of quarks and gluons.
}
\end{figure}

The quark model of hadron classification proposed by 
Gell--Mann and  Zweig in 1964 starts a 15 years--long
term in which sub--hadronic particles, quarks and gluons,
were discovered and a theory of their interactions,
quantum chromodynamics  was established. A brief
history of this term is shown in Fig.~\ref{fig:quarks-gluons}. 
In parallel, conjectures were formulated concerning 
the existence and properties  of matter consisting of sub--hadronic
particles~\cite{Ivanenko:1965dg,Itoh:1970uw}, 
soon called a quark--gluon plasma and studied in detail
within the QCD~\cite{Shuryak:1980tp}. 
The first QCD--inspired estimate of the transition temperature to
QGP gave $T_C \approx 500$~MeV~\cite{Shuryak:1977ut}

Figure~\ref{fig:quarks-gluons} presents 
a brief history of discoveries of quarks and gluons.

Many physicists started to speculate that the QGP can be formed in
nucleus--nucleus collisions at high energies and thus it may
be discovered in laboratory experiments.

\newpage
\section{Statistical QGP hadronization and statistical parton production}

Questions concerning QGP properties and properties of its transition
to matter consisting of hadrons were considered since the late 70s.
\emph{
Cabibbo, Parisi~\cite{Cabibbo:1975ig}, 
Hagedorn and Rafelski~\cite{Hagedorn:1980cv} 
suggested that the 
upper limit of the hadron temperature, the  Hagedorn
temperature, is the transition temperature
to the QGP.  
Furthermore,
Rafelski~\cite{Rafelski:1991rh}
and the collaborators introduced the statistical
approach to the QGP hadronization. It predicted that the resulting
system of hadrons is in incomplete equilibrium.  
The deviations from the equilibrium can be traced back 
to the
QGP properties~\cite{Letessier:2005qe}.
}

In the mid--90s the Statistical Model of the Early Stage (SMES)
was formulated~\cite{Gazdzicki:1998vd}
as an extension of    
Fermi's statistical model of hadron production.
It assumes a statistical production of confined matter at
low collision energies (energy densities) and 
a statistical QGP creation at high collision energies 
(energy densities).
The model predicts a rapid change of the collision energy dependence
of hadron production properties, that are sensitive to QGP, as a
signal of a transition to quark--gluon plasma 
(the onset of deconfinement) in nucleus--nucleus collisions.
The onset energy was estimated to be located in the CERN SPS energy range.

\begin{figure}[!h]
\begin{center}
\includegraphics[width=0.6\columnwidth]{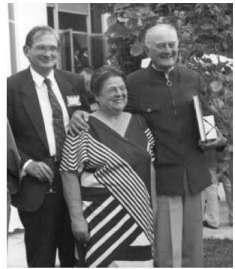}
\end{center}
\vspace*{-0.2cm}
  \caption{\label{fig:statistical-qgp}
Statistical QGP hadronization and statistical parton production.
In photo from left to right are  Johann Rafelski, Tatiana Faberge
(former Cern
theory group Secretary and owner of the rights to Faberge eggs) and
Rolf Hagedorn, Divonne--les--Bains, 1994.
}
\end{figure}

Clearly,
the QGP hypothesis and the SMES model removed the contradiction between
Fermi's and Hagedorn's statistical approaches (see Sect.~5).
Namely, the early stage temperature of strongly interacting matter
is unlimited and increases monotonically with collisions energy,
whereas there is a maximum temperature of hadron gas,
$T_C = T_H \approx 150$~MeV, above which strongly
interacting matter is in a quark--gluon plasma phase.  
Figure~\ref{fig:statistical-qgp} summarizes 
the developments discussed above.

\newpage
\section{QCD--based and QCD--inspired models}

Attempts to derive from the QCD precise quantitative
predictions
for multi--particle production in high energy collisions 
have not been successful~\cite{Brock:1993sz}.
Predictions of QCD--based and QCD--inspired models suffer
from large quantitative  and 
large qualitative  uncertainties, respectively.

The power--law dependence of transverse momentum spectra,
$d\sigma /dp_T \sim 1/p_T^{-4}$,
at high $p_T$ (in the perturbative domain) was
predicted by Field and Feynman in 1977 based on the asymptotic
freedom of the QCD.
In fact, data at $p_T > 2-3$~GeV follow the power--law dependence,
but with the power $-8$~\cite{Affolder:2001fa} instead of $-4$.
Parton cascade and hadronization models were 
developed~\cite{Ellis:1995pe},
however their
validity is in question due to statistical and hydrodynamical features
of experimental data\cite{Florkowski_textbook}, which are difficult to explain  within
the dynamical, QCD--inspired approaches~\cite{Mrowczynski:2006ad}.

\begin{figure}[!h]
\begin{center}
\includegraphics[width=1.0\columnwidth]{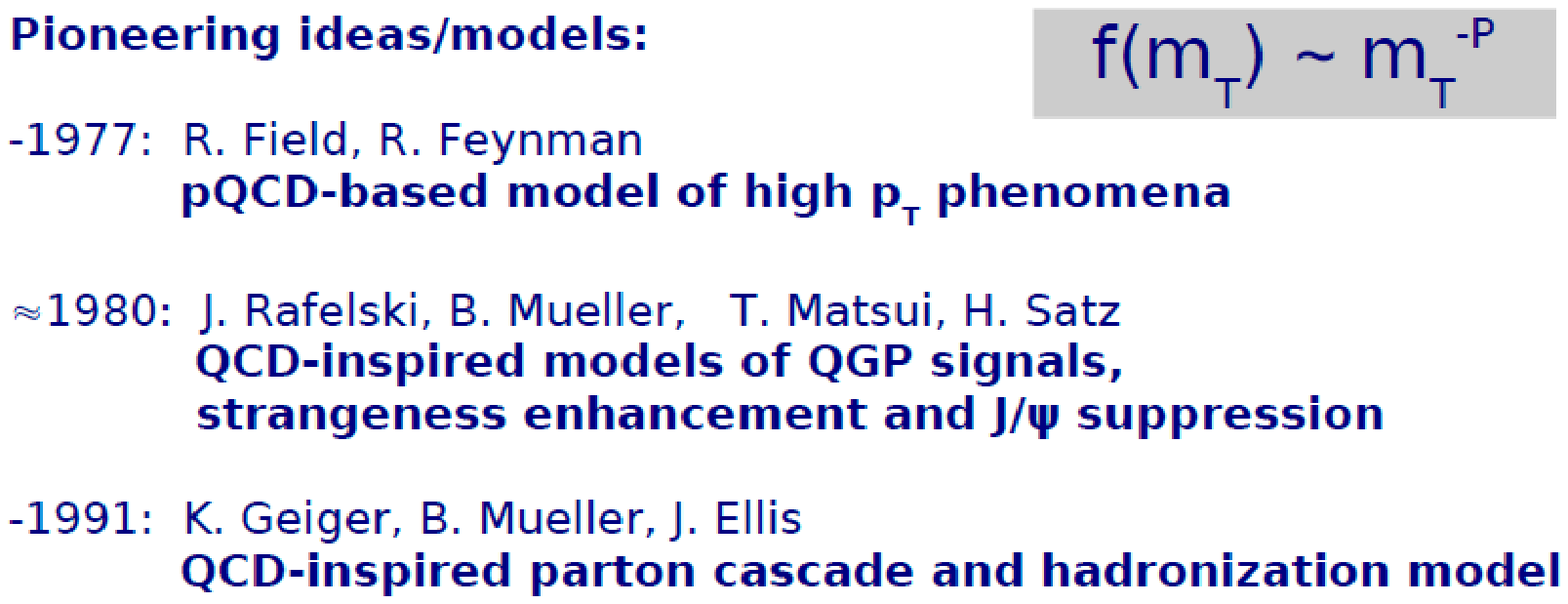}
\end{center}
  \caption{\label{fig:qcd}
Pioneering
QCD--based and QCD--inspired ideas and models.
}
\end{figure}

\emph{
The most famous 
QCD--inspired models of QGP signals in nucleus--nucleus collisions,
strangeness enhancement~\cite{Rafelski:1982pu}
and $J/\psi$ suppression~\cite{Matsui:1986dk}, 
were proposed in 1980s by Rafelski, M\"uller and Matsui, Satz, respectively.
They motivated precision 
systematic measurements.} 
Nevertheless,
they did not lead to definite conclusions on the QGP creation, because
alternative, QCD--inspired and/or statistical explanations of data
were more successful.
Figure~\ref{fig:qcd} summarizes 
the developments discussed above.

\newpage
\section{Discoveries of strongly interacting matter and its phase transition}

Systematic data from experiments at the CERN SPS and LHC and
at the BNL AGS and RHIC clearly indicate that a system of
strongly interacting particles created in
heavy collisions at high energies is close to, at least local,
equilibrium. At freeze-out the system occupies a volume which is
much larger than a volume of  individual hadron. The latter conclusion
is based on the failure of the WNM and the success of 
statistical~\cite{Becattini:2009sc} and
hydrodynamical models~\cite{Florkowski_textbook}.
Thus, one concludes that strongly interacting matter
is created in 
heavy ion collisions. 

The phase transition of strongly interacting matter to the QGP
was discovered within the energy scan program of NA49 at
the CERN SPS~\cite{Afanasiev:2002mx,:2007fe}. 
The program was motivated by the predictions of the SMES model.
The discovery is based on the observation
that several basic hadron production properties measured
in heavy ion collisions rapidly change their dependence on
collisions energy in a common energy 
domain~\cite{Gazdzicki:2010iv}.

Figure~\ref{fig:qcd} summarizes 
the developments discussed above.

\begin{figure}[!h]
\begin{center}
\includegraphics[width=0.9\columnwidth]{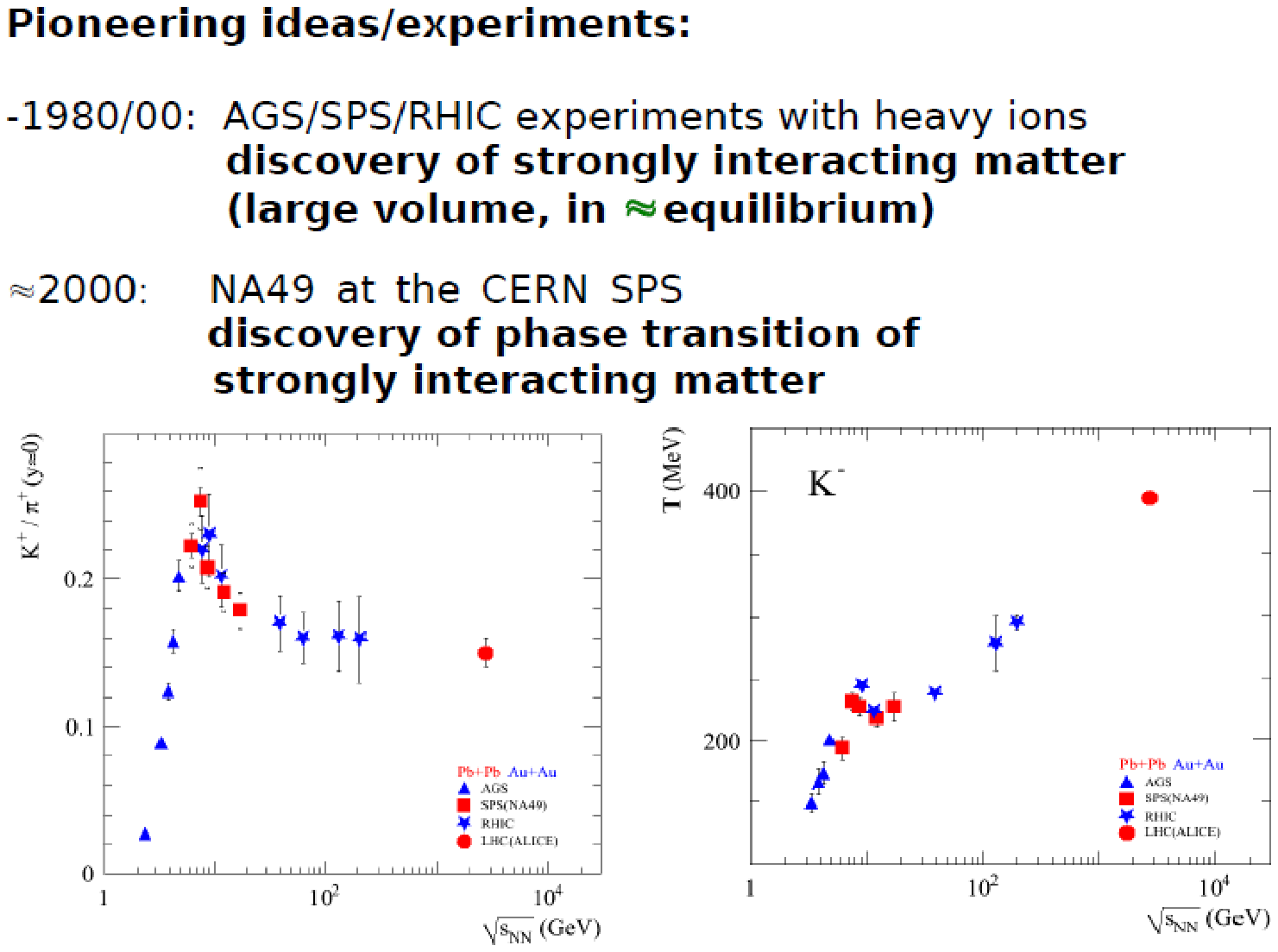}
\end{center}
  \caption{\label{fig:onset}
Discoveries of strongly interacting matter and its phase transition.
The horn (left) and step (right) structures
in energy dependence of the $K^+/\pi^+$ ratio and the inverse slope 
parameter of $K^-$ $m_T$ spectra signal the onset of deconfinement
located at the low CERN SPS energies. 
}
\end{figure}

\newpage

\section{Acknowledgments}
\label{Acknowledgments}
I would like to thank the organizing committee of the SQM 2011 conference 
for the possibility to contribute to the jubilee session dedicated to
the 60th birthday of
Johann Rafelski.

This work was supported by
the German Research Foundation under grant GA 1480/2-1.

\end{document}